\documentclass[12pt]{article}
\usepackage{graphics,psfrag}
\usepackage{amsthm,amssymb,epsfig,amsmath,euscript,array,cite}

\newcounter{multieqs}

\newcommand{\be}{\begin{equation}}
\newcommand{\ee}{\end{equation}}
\newcommand{\eq}[1]{(\ref{#1})}

\newcommand{\bm}[1]{\mbox{\boldmath $#1$}}
\newcommand{\rf}[1]{(\ref{#1})}

\def\bd{\begin{document}}
\def\ed{\end{document}}
\def\nn{\nonumber}
\def\bea{\begin{eqnarray}}
\def\eea{\end{eqnarray}}
\let\bm=\bibitem
\let\la=\label

\def\npb#1#2#3{Nucl. Phys. {\bf{B#1}} #3 (#2)}
\def\plb#1#2#3{Phys. Lett. {\bf{#1B}} #3 (#2)}
\def\prl#1#2#3{Phys. Rev. Lett. {\bf{#1}} #3 (#2)}
\def\prd#1#2#3{Phys. Rev. {D \bf{#1}} #3 (#2)}
\def\cmp#1#2#3{Comm. Math. Phys. {\bf{#1}} #3 (#2)}
\def\cqg#1#2#3{Class. Quantum Grav. {\bf{#1}} #3 (#2)}
\def\nppsa#1#2#3{Nucl. Phys. B (Proc. Suppl.) {\bf{#1A}}#3 (#2)}
\def\ap#1#2#3{Ann. of Phys. {\bf{#1}} #3 (#2)}
\def\ijmp#1#2#3{Int. J. Mod. Phys. {\bf{A#1}} #3 (#2)}
\def\rmp#1#2#3{Rev. Mod. Phys. {\bf{#1}} #3 (#2)}
\def\mpla#1#2#3{Mod. Phys. Lett. {\bf A#1} #3 (#2)}
\def\jhep#1#2#3{J. High Energy Phys. {\bf #1} #3 (#2)}
\def\atmp#1#2#3{Adv. Theor. Math. Phys. {\bf #1} #3 (#2)}

\def\N{{\cal N}}
\def\sst{\scriptscriptstyle}
\def\thetabar{\bar\theta}
\def\Tr{{\rm Tr}}
\def\one{\mbox{1 \kern-.59em {\rm l}}}

\def\a{\alpha}      \def\da{{\dot\alpha}}  
\def\b{\beta}       \def\db{{\dot\beta}}  
\def\g{\gamma}  \def\G{\Gamma}  \def\dc{{\dot\gamma}}  
\def\d{\delta}  \def\D{\Delta}  \def\ddt{\dot\delta}  
\def\e{\epsilon}        \def\vare{\varepsilon}  
\def\f{\phi}    \def\F{\Phi}    \def\vvf{\f}  
\def\h{\eta}  
\def\k{\kappa}  
\def\l{\lambda} \def\L{\Lambda}  
\def\m{\mu} \def\n{\nu}  
\def\o{\omega}  
\def\p{\pi} \def\P{\Pi}  
\def\r{\rho}  
\def\s{\sigma}  \def\S{\Sigma}  
\def\t{\tau}  
\def\th{\theta} \def\Th{\Theta} \def\vth{\vartheta}  
\def\X{\Xeta}  
\def\z{\zeta}  

\def\na{\nabla}  
\def\cA{{\cal A}} \def\cB{{\cal B}} \def\cC{{\cal C}}  
\def\cD{{\cal D}} \def\cE{{\cal E}} \def\cF{{\cal F}}  
\def\cG{{\cal G}} \def\cH{{\cal H}} \def\cI{{\cal I}}  
\def\cJ{{\cal J}} \def\cK{{\cal K}} \def\cL{{\cal L}}  
\def\cM{{\cal M}} \def\cN{{\cal N}} \def\cO{{\cal O}}  
\def\cP{{\cal P}} \def\cQ{{\cal Q}} \def\cR{{\cal R}}  
\def\cS{{\cal S}} \def\cT{{\cal T}} \def\cU{{\cal U}}  
\def\cV{{\cal V}} \def\cW{{\cal W}} \def\cX{{\cal X}}  
\def\cY{{\cal Y}} \def\cZ{{\cal Z}}

\def\ua{\underline{\alpha}}  
\def\ub{\underline{\phantom{\alpha}}\!\!\!\beta}  
\def\uc{\underline{\phantom{\alpha}}\!\!\!\gamma}  
\def\um{\underline{\mu}}  
\def\ud{\underline\delta}  
\def\ue{\underline\epsilon}  
\def\una{\underline a}\def\unA{\underline A}  
\def\unb{\underline b}\def\unB{\underline B}  
\def\unc{\underline c}\def\unC{\underline C}  
\def\und{\underline d}\def\unD{\underline D}  
\def\une{\underline e}\def\unE{\underline E}  
\def\unf{\underline{\phantom{e}}\!\!\!\! f}\def\unF{\underline F}  
\def\unm{\underline m}\def\unM{\underline M}  
\def\unn{\underline n}\def\unN{\underline N}  
\def\unp{\underline{\phantom{a}}\!\!\! p}\def\unP{\underline P}  
\def\unq{\underline{\phantom{a}}\!\!\! q}  
\def\unQ{\underline{\phantom{A}}\!\!\!\! Q}  
\def\unH{\underline{H}}  
  
\def\As {{A \hspace{-6.4pt} \slash}\;}  
\def\bs {{b \hspace{-6.4pt} \slash}\;}  
\def\Ds {{D \hspace{-6.4pt} \slash}\;}  
\def\ds {{\del \hspace{-6.4pt} \slash}\;}  
\def\ss {{\s \hspace{-6.4pt} \slash}\;}  
\def\ks {{ k \hspace{-6.4pt} \slash}\;}  
\def\ps {{p \hspace{-6.4pt} \slash}\;}   
\def\xs {{x \hspace{-6.4pt} \slash}\;}  
\def\pas {{{p_1} \hspace{-6.4pt} \slash}\;}  
\def\pbs {{{p_2} \hspace{-6.4pt} \slash}\;}  
  
\def\Dh{\hat{D}}
\def\Gh{\hat{G}}
\def\Fh{\hat{F}}
\def\Ph{\hat{P}}
\def\Rh{\hat{R}}
\def\Vh{\hat{V}}  
\def\Xh{\hat{X}} 
 
\def\ah{\hat{a}}
\def\gh{\hat{g}} 
\def\hh{\hat{h}}
\def\uh{\hat{u}}  
\def\xh{\hat{x}}  
\def\yh{\hat{y}}  
\def\ph{\hat{p}}  
\def\xih{\hat{\xi}}  
\def\chih{\hat{\chi}}

\def\psit{\tilde{\psi}}  
\def\Psit{\tilde{\Psi}}   
\def\Psibt{\tilde{\bar{Psi}}}  

\def\Phit{\tilde{\Phi}}   
\def\Phitb{\overline{\tilde{Phi}}}  
\def\tht{\tilde{\th}}  
\def\lt{\tilde{\l}}
\def\chit{\tilde{\chi}}   

\def\At{\tilde{A}}
\def\Dt{\tilde{D}}
\def\Ft{\tilde{F}}
\def\Qt{\tilde{Q}}  
\def\Rt{\tilde{R}}  
\def\Mt{\tilde{M }}  
\def\Nt{\tilde{N}}   
\def\Xt{\tilde{X}}
 
\def\at{\tilde{a}}  
\def\htt{\tilde{h}} 
\def\st{\tilde{s}}  
\def\ft{\tilde{f}}
\def\gt{\tilde{g}}
\def\pt{\tilde{p}}  
\def\qt{\tilde{q}}  
\def\vt{\tilde{v}}  
\def\nt{\tilde{n}}  
\def\ut{\tilde{u}} 
\def\xt{\tilde{x}} 
\def\yt{\tilde{y}} 
\def\Psit{\tilde{\Psi}}
\def\vphit{\tilde{\varphi}}  

\def\delb{\bar{\partial}}  
\def\thb{\bar{\theta}}
\def\mub{\bar{\mu}}
\def\lamb{\bar{\l}}
\def\psib{\bar{\psi}}
\def\sb{\bar{\sigma}}
\def\xib{\bar{\xi}}
\def\chib{\bar{\chi}}

\def\Phib{\bar{\Phi}}
\def\Lamb{\bar{\Lambda}}

\def\Ab{{\overline A}} \def\Bb{{\overline B}} \def\Cb{{\overline C}}  
\def\Db{{\overline D}} \def\Eb{{\overline E}} \def\Fb{{\overline F}}  
\def\Gb{{\overline G}} \def\Hb{{\overline H}} \def\Ib{{\overline I}}  
\def\Jb{{\overline J}} \def\Kb{{\overline K}} \def\Lb{{\overline L}}  
\def\Mb{{\overline M}} \def\Nb{{\overline N}} \def\Ob{{\overline O}}  
\def\Pb{{\overline P}} \def\Qb{{\overline Q}} \def\Rb{{\overline R}}  
\def\Sb{{\overline S}} \def\Tb{{\overline T}} \def\Ub{{\overline U}}  
\def\Vb{{\overline V}} \def\Wb{{\overline W}} \def\Xb{{\overline X}}  
\def\Yb{{\overline Y}} \def\Zb{{\overline Z}}  

\def\fb{{\overline f}}
\def\gb{{\overline g}}
\def\mb{{\overline m}}
\def\lb{{\overline l}}
\def\yb{{\overline y}}

\def\ba{{\bf a}} 
\def\bk{{\bf k}}  
\def\bl{{\bf l}}  
\def\bp{{\bf p}}  
\def\bq{{\bf q}}  
\def\br{{\bf r}}
\def\bt{{\bf t}}
\def\bu{{\bf u}}
\def\bv{{\bf v}}
\def\bx{{\bf x}}  
\def\by{{\bf y}}  
\def\bR{{\bf R}}  
\def\bV{{\bf V}}

\def\bone{{\bf 1}}  

\def\va{{\vec a}}
\def\vk{{\vec k}}
\def\vp{{\vec p}}
\def\vq{{\vec q}}
\def\vx{{\vec x}}
\def\vy{{\vec y}}
\def\vu{{\vec u}}
\def\vv{{\vec v}}

\def\vs{{\vec \sigma}}
\def\vtau{{\vec \tau}}

\newcommand{\ov}[1]{\overrightarrow{#1}}
  
\def\d{\delta}\def\D{\Delta}\def\ddt{\dot\delta}  
  
\def\pa{\partial} \def\del{\partial}  
\def\xx{\times}  
\def\uno{\mbox{1 \kern-.59em {\rm l}}}    
  
\def\trp{^{\top}}  
\def\inv{^{-1}}  
\def\dag{{^{\dagger}}}  
\def\pr{^{\prime}}  
  
\def\rar{\rightarrow}  
\def\lar{\leftarrow}  
\def\lrar{\leftrightarrow}  
  
\newcommand{\0}{\,\!}      %this is just NOTHING!  
\def\one{1\!\!1\,\,}  
\def\im{\imath}  
\def\jm{\jmath}  
  
\newcommand{\tr}{\mbox{tr}}  
\newcommand{\slsh}[1]{/ \!\!\!\! #1}  
  
\def\vac{|0\rangle}  
\def\lvac{\langle 0|}  
  
\def\hlf{\frac{1}{2}}  
\def\ove#1{\frac{1}{#1}}  

\def\Box{\square}  
\def\ZZ{\mathbb{Z}}  
\def\bb#1{{\bf #1}}  
\def\bcomment#1{}  
\def\bfhat#1{{\bf \hat{#1}}}  
\def\VEV#1{\left\langle #1\right\rangle}  

\newcommand{\ex}[1]{{\rm e}^{#1}} \def\ii{{\rm i}}  

\newcommand{\lrbrk}[1]{\left(#1\right)}
\newcommand{\sfrac}[2]{{\textstyle\frac{#1}{#2}}}
 
\def\stw{{\sqrt{2}}}

\def\rf {{\rm f}}
\def\ri {{\rm i}}
\def\rs {{\scriptscriptstyle \rm S}}
\def\rt {{\scriptscriptstyle \rm T}}

\def\rQ {{\scriptscriptstyle \rm \cQ}}
\def\rR {{\scriptscriptstyle \rm \cR}}

\def\cQb{{\cal \Qb}}
\def\cRb{{\cal \Rb}}
\def\cWb{{\cal \Wb}}

\def\fd {{\rm N}}
\def\afd {{\overline{\rm N}}}

\def \II {I\hspace{-.1em}I\hspace{.1em}}
\def \IIA {\mbox{\II A\hspace{.2em}}}
\def \IIB {\mbox{\II B\hspace{.2em}}}
\def \gs {g^s}
\def \ls {\lambda^s}

\def \I {{\cal I}}
\def \qs {q\hspace{-.53em}/\hspace{.15em}}
\def \ks {k\hspace{-.53em}/\hspace{.15em}}
\def \YM {{\mbox{\tiny YM}}}
\def \gym {g_{\YM}}

\def \Lc {\L_c}

\begin{document}
  
\hfill{arXiv:0710.2640}

\vspace{20pt}

\begin{center}

{\Large \bf  Time-dependent AdS/CFT Duality \II: \\
Holographic Reconstruction of Bulk Metric \\
and Possible Resolution of Singularity 
%from the Wilsonian Effective Action 
}
\vspace{30pt}

{\bf Chong-Sun Chu$^1$, Pei-Ming Ho$^2$}

\vspace{15pt}
{\em
\begin{itemize}
\item[$^1$]
Centre for Particle Theory and Department of Mathematics,
Durham University, \\Durham, DH1 3LE, UK.
\item[$^2$]
Department of Physics and Center for Theoretical Sciences, 
National Taiwan University, Taipei 10617, Taiwan, 
R.O.C.
\end{itemize}
}

\vskip .1in {\small \sffamily chong-sun.chu@durham.ac.uk, 
pmho@phys.ntu.edu.tw}

\vspace{50pt}
{\bf Abstract}
\end{center}

We continue the studies of 
our earlier proposal 
for an AdS/CFT correspondence for 
time-dependent supergravity backgrounds.
We note that by performing a suitable change of variables, 
the dual super Yang-Mills theory 
lives on a flat base space, and the time-dependence of 
the supergravity background is 
entirely encoded in 
the time-dependent couplings (gauge and axionic) and 
their supersymmetric completion. This form of the SYM allows 
a detailed 
perturbative
analysis to be performed.
In particular the 
one-loop 
Wilsonian effective action of the boundary SYM theory 
is computed. 
By using the holographic UV/IR relation, 
we propose a way to extract the bulk metric from the Wilsonian effective
action; and we
find that the bulk metric of our supergravity solutions can be
reproduced precisely. 
While the bulk geometry can have various singularities such as
geodesic incompleteness, gauge theory 
quantum
effects can introduce 
higher derivative corrections in the effective action  which
can serve as a way to resolve the singularities.

\setcounter{page}0
\newpage

\section{Introduction}

The understanding of the nature of spacetime singularity, 
and whether and how it is resolved, is one of the most important question for a
quantum theory of gravity. 
Recently, powerful nonperturbative formulations of string theory such
as Matrix theory \cite{mat, mat-review} and AdS/CFT correspondence 
\cite{ads1,ads2,ads3,mal-review} have been put forward and 
%c8 much studied.
intensively studied for various applications.
While much work has been devoted to the studies of blackhole
singularities \cite{bh}, there were much fewer studies on spacetime
singularities of cosmological type.
It is desirable to apply these ideas to the studies of time-dependent
backgrounds, and try to use them to learn about spacetime singularity.
See \cite{sing-review} for recent reviews on approaches to understanding
spacelike or null singularities in string theory.

In \cite{CH}, we constructed a supersymmetric AdS/CFT correspondence
for
a class of 
time-dependent \IIB backgrounds.
The supergravity (SUGRA) backgrounds have nontrivial time 
dependence through a null coordinate. 
Similar SUGRA backgrounds were also constructed 
in \cite{lin,d1}. 
In addition we have also constructed the dual gauge theory explicitly \cite{CH}. 
The gauge theory features 
a time-dependent gauge coupling  and a time-dependent axion coupling.
The proposed gauge/gravity duality thus 
constitutes
a natural 
starting point for understanding time-dependent superstring backgrounds 
from the super Yang-Mills (SYM) theory. 
Our work was motivated by the
earlier works of \cite{HH} which 
proposed
to use AdS/CFT 
correspondence
to study a big crunch cosmology; and \cite{csv} which 
proposed
to use matrix string theory to study null singularities.
%c10 % p5be \cite{Craps2006,LiSong,LinTomino}.
For a different approach of applying AdS/CFT
to time-dependent backgrounds see \cite{Cvetic:2003zy}.
% p5ee
See also \cite{related1} for related works.

In our construction \cite{CH}, the SUGRA metric admits cosmological, 
null-like singularities for some class of dilaton and axion 
field configurations.  
Moreover, since these singularities are situated at a constant $x^+$ 
(rather than localized at finite radial coordinate),
their presence can in principle be detected 
by quantities computed in the dual field theory. In \cite{CH},
we have carried out a generic analysis at the free field level.  We found that 
the field theory two-point functions computed from the gravity side using 
the duality is
different from the one computed directly from the field
theory, which up to a
rescaling of fields, is the same as the one defined for an ordinary
Minkowskian spacetime. 
In particular, the SUGRA
result is sensitive to the singularity of the spacetime, while the
gauge theory result does not see the singularity. 
That the results differ is not surprising since the SUGRA result 
is valid in the regime where the t'Hooft
coupling is  large, while the field theory result 
is valid when the t'Hooft coupling is small. 
We interpreted our results as suggesting that the spacetime singularity seen at
the SUGRA level could be resolved by $\alpha'$ effects of string theory.
Similar analysis has also been performed in \cite{d2} 
with the same conclusion.

While this might look encouraging, the following remarks prompt immediately
for more detailed studies in the dynamics of the gauge theory. First,
the regularity of the field theory 2-point function is demonstrated only 
at the 
free field level.
When interaction is included, we expect the answer to depend on the coupling,
as well as its higher time derivatives. It is therefore possible that
the field theory result 
%c8 becomes 
could reproduce the SUGRA singularity once
%c8 once the higher loop
quantum 
corrections are included. 
Secondly, in our work
\cite{CH}, 
we pointed out that
the Einstein equation (see \eq{eom} below), 
which constitutes
a constraint on the dilaton and axion field, 
can be obtained from the requirement of
finiteness of the energy momentum tensor 
in gauge theory.
However, this argument is not completely satisfactory as it is based on
the validity of the duality. In \cite{CH}, we conjectured that 
the Einstein equation could be derived from the
the gauge theory at the quantum level. 
Achieving this would help us to understand better how 
%c7 generic 
quantum
properties of SYM is mapped holographically to the 
geometrical
properties of spacetime.

The purpose of this paper is to go beyond the free field level analysis 
and to try to understand
the role of the SYM quantum effects in the duality. In
particular we 
propose to identify bulk 
metric
properties from the Wilsonian effective action obtained by
intergrating out the high momentum modes. 

In section 2 we review our
proposal of the time-dependent AdS/CFT duality. In section 3, we demonstrate that  
by a change of variables, both the SUGRA metric and the SYM Lagrangian can be
written in a simpler form. We restate our duality proposal in this
frame. In section 4, we present the Feynman rules for our time-dependent
SYM theory. The presence of the time-dependent gauge couplings and time-dependent 
theta angle modify the interaction vertices.
In section
5, we compute the 1-loop Wilsonian fermion kinetic term and find that at
the leading order of derivative expansion, it allows one to reconstruct 
the bulk metric of the gravity side.
It is straightforward to include the higher order derivative corrections
to the holographically constructed bulk geometry
and we discuss how spacetime singularity could be resolved.

\section{Time-dependent AdS/CFT correspondence: review}

In \cite{CH},   a time-dependent deformation to the original 
AdS/CFT correspondence was constructed. The non-vanishing fields consist of the 
Einstein metric
\begin{equation} \label{sugra-metric-rosen}
ds^2 = \frac{R^2}{u^2} \left(
- k^2(x^+)dx^+ dx^-  + M_i^2(x^+) (d x^i)^2 + du^2 \right)
+ R^2 d\Omega_5^2 \qquad (i = 2,3),  
\end{equation} 
an undeformed 5-form, and dilaton and
axion fields $\phi(x^+), \chi(x^+)$. All equations of motion are satisfied provided  
\begin{equation} \label{eom}
\frac{1}{2} (\phi')^2 + \frac{1}{2} e^{2\phi}(\chi')^2  
= -\sum_{i=2,3} 
\left( \frac{M_i''}{M_i} - \frac{2k'M_i'}{k M_i} \right),
\end{equation}
which 
comes
from the $(++)$-component of the Einstein equation.

The SUGRA solution preserves eight \IIB supersymmetries. 
Viewed as a
deformation of the standard $AdS_5 \times S^5$, half of the Poincare
supersymmetry is preserved, %all the "AdS-supersymmetries" are broken.
and the conformal supersymmetry is broken.
The solution can be
obtained from a near horizon limit of a strack of D3 branes with a
pp-wave on it. This gives rise to the relation between the radius $R$
and the dilaton \cite{CH}
\begin{equation} \label{RR}
R^4 = 16 \pi N \langle g_s^{-1} \rangle^{-1} l_s^4 ,
\end{equation}
where
$\langle g_s^{-1} \rangle := \int dx^+ k^2 e^{-\phi} /
\int dx^+ k^2$
is the $x^+$-average of the inverse of the string coupling
$g_s = e^{\phi}$.  This relation generalizes the celebrated relation in
the original AdS/CFT correspondence, and is a consequence of the BPS property of 
the stack of pp-wave D3-branes. 
It is interesting to note that if $\langle g_s^{-1} \rangle$ diverges, 
which could happen if $g_s$ goes to zero somewhere, 
$N$ needs to be infinity even for finite radius $R$.

We also noted that the supergravity solution  
is invariant under the  scaling transformation 
\begin{equation} \label{scaling-sugra}
u \to \lambda u, \quad x^+ \to x^+, \quad x^- \to \lambda^2 x^-, 
\quad x^i \to \lambda x^i.
\end{equation}
The same symmetry is respected by our time-dependent SYM \cite{CH}.

The above metric was written down in the Rosen form. 
One can also perform a change of coordinate \cite{CH} 
to put the metric in the following
Brinkman form 
\be
ds^2 =\frac{R^2}{u^2} \left(
- k^2(x^+)dx^+ dx^-  + h(x^+, x^i) (dx^+)^2
+ (d x^i)^2 + du^2 \right), \label{brinkman1} 
\ee
where
\be
h(x^+, x^i) = \sum_{i=2,3} h_i(x^+) \;(x^i)^2, \quad \mbox{and}\quad
h_i(x^+)= \frac{M''_i }{M_i} -  \frac{2k' M_i'}{k M_i } .
\ee

% p5be
Without loss of generality, one can choose the coordinate $x^+$ 
such that $k(x^+) = 1$.
It is then easy to see that this metric has an interesting property, 
that is, 
it deviates from the undeformed $AdS$ metric only 
by the $g_{++}$ (or equivalently the $g^{--}$) component, 
and as a result only the $R_{++}$ component 
of the curvature tensor is modified.
It follows that all the invariants obtained by
contracting indices of curvature tensors 
are exactly the same as pure $AdS$ space.
Therefore we believe that this metric is free from 
stringy ($\alpha'$) corrections for the same reason 
why $AdS$ is an exact consistent background.
On the other hand, our background is expected to receive 
$g_s$ corrections due to string loop diagrams.
% p5ee

Provided that the radius $R$ defined by \eq{RR} is well defined, 
we proposed in \cite{CH} that
the quantum gravity for the time-dependent background \eq{brinkman1}
is dual to a  SYM theory living on the boundary metric
\be \label{ym-metric-h}
ds_\YM^2 =
- k^2(x^+)dx^+ dx^-  + h(x^+, x^i) (dx^+)^2
+ (d x^i)^2
\ee
and has a time-dependent Yang-Mills coupling  
and theta angle
\be \label{iden1}
\frac{\th}{2\pi} + \frac{4\pi i}{\gym^2}=\chi + i e^{-\phi}.
\ee
The Lagrangian density is  
\footnote{There was 
a typo in the fermionic part of the axion action $S_\chi$ in our previous paper
\cite{CH}. $\chi$ should be replaced by $\chi'$ as in \eq{S4}.} 
$\cL = \cL_\YM + \cL_X + \cL_\Psi +\cL_\chi$,
where \cite{CH}
\bea
&& \cL_{\YM} =  \frac{\sqrt{-g}}{\gym^2}
\Tr \left( - \frac{1}{4} g^{\m\m'} g ^{\n\n'}
F_{\m\n} F_{\m'\n'}  \right),  \label{S1}\\
&& \cL_X = \frac{\sqrt{-g}}{\gym^2} \Tr \left(
-\frac{1}{2}g^{\m\n} D_\m X^a  D_\n X^a
+\frac{1}{4}[X^a,X^b]^2 \right), \label{S2}\\
&& \cL_\Psi = 
\frac{\sqrt{-g}}{ \gym^2} \Tr \left( 
\frac{1}{2} \bar{\Psi} \gamma^\m [ -i D_\m, \Psi] 
+ \frac{1}{2} \bar{\Psi} \gamma^a [ X_a, \Psi] \right), \label{S3} \\
&& \cL_{\chi} = 
 \frac{1}{8\pi^2}\mbox{Tr} \left( 
- \frac{1}{4} \th(x^+) \epsilon^{\m\n\r\s} F_{\m\n}
 F_{\r\s}
 + 
\frac{i}{4}
\th'(x^+) \bar{\Psi}\Gamma^2\Gamma^3\Gamma^+\Psi \right).
 \label{S4}
\eea
Here $\Psi$ is a Majorana spinor, $D_\mu = \del_\mu + i A_\mu$ and 
$F_{\m\n} = \del_\m A_\n -\del_\n A_\m + i [A_\m,A_\n]$
($\m,\n =+,-,2,3$; $a=4,\cdots,9$).
The spin connection term  vanishes since
the only nonvanishing components of the spin connection are
\be
\o_{i+} = \frac{1}{2}(\del_i h) E^+, \quad \o_{-+} = \frac{k'}{k} E^+ , 
\ee
where $E^A$ denotes the vielbein, 
and so
\be 
\bar{\Psi} \gamma^\mu \o_{\mu AB} \G^{AB}\Psi \sim 
\bar{\Psi} \G^{-} \Psi =0, 
\ee
where we have used the fact that
$ \Gamma^0 \Gamma^{\mu}$ is symmetric
for Majorana representation.

On the gravity side, the SUGRA solution
is invariant under 8 supersymmetries
satisfying
\begin{equation}
\Gamma^+ \e =  0, \quad 
(1- \Gamma^r) \e =0,
\label{g-eps}
\end{equation} 
where the $r$-direction is defined from substituting $u=e^r$ in the metric.
On the gauge theory side, we have
$\Gamma^+ \e = 0$ and the
usual conformal SUSY transformation with 
$\epsilon = x^\mu \Gamma_\mu \eta$ for ${\cal N}=4$ SYM is broken.
We have also normalized $\cL_{\chi}$ so that when $\th' =0$
it reduces to the standard $\th$-Lagrangian, i.e., it is $\th$ times 
the instanton number.

An interesting feature of our supergravity solutions is that they can admit
singularity. This happens when the Ricci curvature component $R_{++}$
becomes singular, in which case the null geodesic along $x^+$ cannot be
extended
beyond the place where $R_{++}$ blows up. 
Using the Einstein equation, this happens when
the scalar field combination
\be
\frac{1}{2}(\phi')^2 +\frac{1}{2}e^{2\phi}(\chi')^2
\ee
diverges. 
Although the classical background may be singular, 
the SYM theory appears to be well defined 
and provides a non-perturbative definition of the quantum gravity theory. 
The singularity is lightlike. This is different from the spacelike 
singularity which 
occurs for the standard big bang and blackhole. Nevertheless,
the understanding of the nature and possible resolution 
of a null singularity is still of great interest.

\section{AdS/CFT duality in simplifying variables}

\subsection{Simplifying the SYM by change of variables}

In \cite{CH} we have performed a free field theory analysis of the duality
written with respect to the frame \eq{sugra-metric-rosen}. 
The two-point function was computed  for the case
$M_2 = M_3 = M$.
 The field theory result was found to be completely regular. In
fact, apart from a rescaling of the field, the two-point function takes
exactly the same form as for a theory defined on a Minkowski space
\cite{CH}, see also \cite{d2}. 
This indeed has a simple explanation.
In this section, we will explain the origin of this scaling and 
show that the SYM theory on this time-dependent 
background is exactly the same as a SYM theory on the flat background 
with a time-dependent coupling.

%c7  In this paper we will also focus on the case $M_2 = M_3 = M$.
Let us consider this case with $M_2 = M_3 = M$.
Without loss of generality, we can choose the coordinate $x^+$ such that 
$k(x^+) =\sqrt{2} M(x^+)$, so
\be
ds^2 = - 2 M^2 dx^+ dx^- + M^2 dx_i^2. 
\ee
For this metric, 
\be
\sqrt{-g} = M^4, \qquad g_{\mu\nu} = M^2 \eta_{\mu\nu}. 
\ee
Let us look at the SYM action term by term. 
First, the YM term \eq{S1} becomes
\be \label{S-YM}
S_{\rm YM} 
= \int d^4 x \frac{1}{\gym^2} \mbox{Tr} \left(
  -\frac{1}{4}\eta^{\mu\nu}\eta^{\a\b}F_{\mu\a}F_{\nu\b} \right),
\ee 
i.e. one which is defined on a flat metric $\eta_{\m\n}$.
This has also been noted by \cite{d2}.
In the following we show that the same is true for  the scalar and 
the fermion action.
Motivated by the
above mentioned rescaling of scalar fields,  
we introduce the rescaled fields 
\be \label{XPsi-rescale}
X^a = M^{-1} Y^a, \quad \Psi = M^{-3/2} \psi.
\ee
The scalar action \eq{S2} becomes
\bea \label{S-X}
S_X 
&=& \int d^4 x \frac{1}{\gym^2} \mbox{Tr} \Biggl[ \eta^{\mu\nu} \left[
-\frac{1}{2} D_{\mu} Y^a D_{\nu} Y^a - \frac{1}{2} \left(\gym^2
  \del_{\nu}\frac{\del_{\mu} M}{\gym^2 M}  
- \frac{\del_{\mu} M}{M}\frac{\del_{\nu}M}{M} \right) {Y^a}^2 \right]
+ \frac{1}{4} [Y^a, Y^b]^2 \Biggr] \nn\\
&=& \int d^4 x \frac{1}{\gym^2} \mbox{Tr}  \left[-\frac{1}{2}
  \eta^{\mu\nu} D_{\mu} Y^a D_{\nu} Y^a 
+ \frac{1}{4} [Y^a, Y^b]^2 \right],
\eea
where in the last line we used the fact that $M(x^+)$ and
$\gym(x^+)$ only depend on $x^+$.  
The fermion action
\footnote{With the choice of vielbein $E^+ = M^2 dx^+, E^- = dx^-, E^i
  = M dx^i$, the nonvanishing component of the spin connection is
  $\o^i_+ =M'/M^3 E^i$. 
We note again the spin connection term in the fermion KE term is zero.} 
becomes
\bea \label{S-Psi}
S_{\Psi}
&=& \int d^4 x \, 
\frac{1}
{\gym^2 } \mbox{Tr}\biggl[ \bar{\psi}
  \Gamma^{\mu} [-iD_{\mu}, \psi]  
+ \frac{3i}{2} \frac{\del_{\mu} M}{M} \bar{\psi} \Gamma^{\mu} \psi
+ \bar{\psi} \Gamma^a [ Y^a, \psi] \biggr] \nn \\
&=& \int d^4 x \, 
\frac{1}
{\gym^2 } \mbox{Tr}\biggl[ \bar{\psi}
  \Gamma^{\mu} [-iD_{\mu}, \psi]  + \bar{\psi} \Gamma^a [ Y^a, \psi] \biggr], 
\eea
where we have introduced the flat space  Gamma matrices $\Gamma^A$. It is
related to the curved space ones $\g^A$ by
\be
\gamma^A = M^{-1} \Gamma^A.
\ee

Thus we see that, remarkably, 
in terms of the new field variables $(A_\m, Y^a, \psi)$,
the dual SYM is defined on a flat base space! 
The curved metric of the bulk simply drops out. 
The only difference from the ordinary $\cN=4$ SYM is the presence of 
the 
time-dependent gauge couplings $\gym$ and $\chi$. 
This difference does not appear at the tree
level, and explains why the field theory two-point functions coincide
with the usual expressions when properly rescaled fields are used to
express the Green's functions \cite{CH}. In addition, our analysis here 
implies that this
is true for a general $n$-point function at the tree level. 
This result for the general $n$-point function 
was first obtained in \cite{d2} using a path-integral argument. Here we
see that both the choice of the rescaled variables and the fact that
the free field theory Green's function is the same as the
Minkowski one 
%p2 both -- the word already used in this sentence
have a very simple explanation.

From the viewpoint of SYM, 
if the coupling $\gym$ approaches to zero at a certain point, 
it only implies that the theory is almost free in the neighborhood of that point.
But from the viewpoint of the bulk supergravity, 
this could correspond to a singularity (e.g. geodesic incompleteness)
in the bulk metric \cite{CH}. 
Understanding the holographic duality of this situation shall lead us to 
a deeper understanding of %c1
the nature of spacetime singularity.

\subsection{Simplifying the bulk metric by change of variables}

The fact that there exists a choice of variables where the SYM theory
takes on a simpler form 
suggests that the same must be true also for
the SUGRA side. We will demonstrate that this is indeed the case now.
 
The metric for the deformed $AdS$ part of the bulk is
\be \label{ds-M}
ds ^2 = \frac{R^2}{u^2}(-2M^2(x^+)dx^+ dx^- + M^2(x^+) dx_i^2 + du^2), 
\ee
where $i = 2,3$. 
Introduce the coordinate change
\bea 
\hat{u} &:=& u/M(x^+), \label{u-rescale}\\
\hat{x}^- &:=& x^- - \frac{1}{2}M^{-1} M' \hat{u}^2,  \label{xm-rescale}
\eea
and then it is easy to show that the metric can be brought to a
Brinkman form, 
\be \label{brinkman2}
ds^2
= \frac{R^2}{\hat{u}^2} \left( -2dx^+ d\hat{x}^- + \frac{1}{2} \Omega
\hat{u}^2 {dx^+}^2 + dx_i^2 + d\hat{u}^2 \right),  
\ee
where $\Omega$ is related to $M$ of \eq{ds-M} and to the dilaton-axion
fields as
\be
\Omega := -2\left( \frac{M''}{M} - 2 \frac{M'^2}{M^2} \right) = 
\frac{1}{2} (\phi')^2 + \frac{1}{2} e^{2 \phi} (\chi')^2.
\ee
The Ricci tensor for \eq{brinkman2} is
\be
\hat{R}_{MN} = - \frac{4 \hat{g}_{MN}}{R^2} + \Omega \d_{M+} \d_{N+}.
\ee

The distinguished feature of the coordinate system \eq{brinkman2} is 
that the boundary
metric at $\hat{u}=0$  
is exactly the same as the boundary of the undeformed $AdS$. In this
frame, the dual SYM theory is deformed only by the presence of
nontrivial time-dependent couplings. It is  given by \eq{S-YM}, \eq{S-X}, 
\eq{S-Psi} plus 
the axionic terms. Moreover, 
the coordinate transformation \eq{u-rescale} (and \eq{xm-rescale}) 
matches the field rescaling  \eq{XPsi-rescale}.

We remark that the metrics \eq{brinkman1} and \eq{brinkman2} are special
cases of the Brinkman form of the metric
\be
ds^2
= \frac{R^2}{u^2} \left( -k^2(x^+) dx^+ d x^- + h(x^+,x^i,u) 
{dx^+}^2 + (dx^i)^2 + d u^2 \right). 
\ee
It has the Ricci tensor 
\be
R_{MN} = \frac{ -4 g_{MN}}{R^2} + \Delta \d_{M+} \d_{N+},  
\ee
where
\be
\Delta :=  \frac{3\del_u h}{2 u} 
-\frac{1}{2}(\del_u^2 h +\del_2^2 h+\del_3^2 h).
\ee
And it is $h =h_{ij}(x^+) x^i x^j$ for \eq{brinkman1} and $h= \Omega(x^+)u^2/2$ 
for \eq{brinkman2}. Einstein equation implies
\be
\Delta = \frac{1}{2} (\phi')^2 + \frac{1}{2} e^{2 \phi} (\chi')^2.
\ee
In the above, we have started with the AdS/CFT duality \cite{CH} 
expressed in the
frame \eq{brinkman1}, and via a series of coordinate transformations, 
related it to the duality expressed in the frame
\eq{sugra-metric-rosen}, and eventually to the duality  expressed in the
frame \eq{brinkman2}.
We could have written down the duality in the frame \eq{brinkman2}
directly. By following the coordinate transformation closely, we have seen how 
a change of coordinate in the bulk corresponds to a (local) 
field re-definition in the dual SYM theory. It is possible that for more
general spacetime diffeomorphism, a non-local field
redefinition in the SYM theory is required.  
It will be  interesting to understand this aspect better. 
We will analysis the duality expressed in the frame
\eq{brinkman2} in the rest of the paper.
%c9 
And we will not write the hat $\hat{}$ over the coordiantes anymore. The
metric of the SUGRA background reads
\be \label{brinkman3}
ds^2
= \frac{R^2}{u^2} \left( -2dx^+ dx^- + \frac{1}{2} \Omega
u^2 {dx^+}^2 + dx_i^2 + du^2 \right). 
\ee

\subsection{Supersymmetric Yang-Mills action}
 
Let us now spell out explicitly the supersymmetry properties of the
SYM theory that is dual to the supergravity solution written in the 
%c9 
frame \eq{brinkman3}
\footnote{ 
The SUSY transformation rule in \cite{CH}  
was only written down for the case of a constant
$\chi$. For the general case where $\chi' \neq 0$, one has to use a
$x^+$ dependent supersymmetry parameter $\e(x^+)$. 
See\eq{ex+} below. }.
Consider the Lagrangian density
\bea
{\cal L}_B &=& %c6 {\cal L}_{\rm YM} + {\cal L}_X =
\frac{1}{4 \gym^2}\mbox{Tr}\left( [Y_M, Y_N][Y^M, Y^N]\right), \label{s1}\\
{\cal L}_{\Psi} &=& 
\frac{1}
{2 \gym^2}\mbox{Tr}\left(
\bar{\psi}\Gamma^M [Y_M, \psi]\right),\\
{\cal L}_{\chi B} &=& \chih(x^+)\mbox{Tr}\left(
\frac{1}{4} \varepsilon^{\mu\nu\a\b} [Y_\m,Y_\n][Y_\a,Y_\b] \right), \\
{\cal L}_{\chi F} &=& 
 \chih'(x^+) \mbox{Tr}\left(
\frac{i}
{4} \bar{\psi}\Gamma^2\Gamma^3\Gamma^+\psi\right),\label{s4}
\eea 
for the fields 
$Y_{\mu} \equiv -i D_{\mu} = -i \del_\m + A_\m, Y^a$, and $\psi$ $(\mu =0,1,2,3;
a=4,\cdots, 9)$.
Here $\varepsilon^{\mu\nu\a\b}$ is the totally antisymmetrized tensor
with $\varepsilon^{+-23} = \varepsilon_{0123} = 1$. 
The metric and  $\Gamma$-matrices are
the ordinary Minkowski ones, $g_{\m\n} = \eta_{\m\n}$ and $g_{ab} = \d_{ab}$. 
For convenience, we have  introduced
the shorthand definition
\be
\chih(x^+) := \frac{\th(x^+)}{8 \pi^2}= \frac{\chi(x^+)}{4\pi}.
\ee
Consider the SUSY transformation 
defined as
\bea
\d Y_{\m} &=&  \d A_{\m} =
-i 
\bar{\epsilon} \Gamma_{\m} \psi, \qquad
\m=0,1,2,3, \label{susy1}\\
\d Y^a &=& 
-i 
\bar{\epsilon} \Gamma^a \psi, \qquad a = 4, \cdots, 9, 
\label{susy2}\\
\d \psi &=& \frac{i}{2} [Y_M, Y_N] \Gamma^{MN}\epsilon, \qquad M, N =
0,1,\cdots, 9, \label{susy3}
\eea
where $\e =\e(x^+)$. 
We have
\be
\d{\cal L}_B = \frac{1}{\gym^2}\mbox{Tr}\left([\d Y_M, Y_N][Y^M, Y^N]\right).
\ee
It is easy to show that
\bea
\d{\cal L}_{\Psi} \simeq -\d {\cal L}_B + \mbox{Tr}
\left(
-\frac{1}
{4}\left(\frac{1}{\gym^2}\right)'
\bar{\epsilon}
\Gamma^{MN}\Gamma^+
\psi [Y_M,Y_N] \right.
&+& 
 \left(\frac{\bar{\epsilon}}{2 \gym^2}\right)' 
\G^{MN}\G^+[Y_M,Y_N]\psi\nn\\
&+&\left.
\left(\frac{1}{\gym^2}\right)' [Y^+,Y^M] \bar{\epsilon}\Gamma_M
\psi
\right),
\eea
where  in this subsection $\simeq$  means equal up to total derivatives.
Assuming that
\be
\Gamma^+\epsilon = 0, \label{assume}
\ee
and using $\G^{MN} \G^+ = \G^+ \G^{MN} + 2 \eta^{+N}\G^M - 2 \eta^{+M}\G^N$,
we have
\be
\d{\cal L}_B + \d{\cal L}_{\Psi} \simeq
\frac{2
\bar{\epsilon}'}{\gym^2} \mbox{Tr} 
\left(\Gamma^M \psi [Y_M,Y^+] \right).
\ee
Thus if 
%c8 the axion field vanishes 
$\chi = 0$,
we can take $\epsilon$ to be constant
and the total action is SUSY invariant.

For 
%c8 a non-zero axion field, 
general  $\chi$, 
the SUSY transformations of the axion
Lagrangian densities are
\bea
\d{\cal L}_{\chi B} &=&  
2
\chih'(x^+) \mbox{Tr}\left(
\varepsilon^{+- j k}\bar{\epsilon}\Gamma_{j}\psi [Y_k,Y^+]\right)
,\qquad j,k =2,3, \\
\d{\cal L}_{\chi F} &=& 
%c8 typo before
%c8 \frac{i}{2}
\chih'(x^+)
\mbox{Tr}\left(
\bar{\epsilon}\Gamma^M\Gamma^2\Gamma^3\psi [Y_M,Y^+ ] \right),
\eea
where we assumed in the derivation that (\ref{assume}) holds.
Therefore, if we choose the transformation parameter to satisfy
\be \label{de}
\frac{2}
{\gym^2}\epsilon' = -
\chih'(x^+) \Gamma^2\Gamma^3\epsilon,
\ee
then
\be
\d({\cal L}_B+{\cal L}_{\Psi}+{\cal L}_{\chi B}+{\cal L}_{\chi F})
=  
- 
\chih'(x^+)\mbox{Tr}\left(
[\Gamma^2\Gamma^3, \Gamma^M]\psi [Y_M,Y^+] -
2 \varepsilon^{+-jk} \bar{\epsilon}\Gamma_{j}\psi [Y_k,Y^+]\right) =0,
\ee
where we have used $[\Gamma^2\Gamma^3, \Gamma^M]= \d^M_i  \cdot
(-2) 
\varepsilon^{+-ij}
 \Gamma^j $ ($i,j =2,3$ here).
The equation \eq{de} is compatible with \eq{assume} and is  solved by
\be
\label{ex+}
\epsilon(x^+) = \exp\left(
{\frac{1}
{4}
\int_0^{x^+}dy^+
\chih'(y^+)\gym^2(y^+)\Gamma_{ch}}\right)
 \epsilon_0, 
\ee
for a constant spinor $\e_0: \Gamma^+ \e_0 =0$. 
Here
$\Gamma_{ch} := \Gamma^+\Gamma^-\Gamma^2\Gamma^3$ 
is the chirality operator and we have used the identity 
$\Gamma^+\Gamma^-\epsilon_0 = -2\epsilon_0$ 
to simplify the expression.

\section{Quantum supersymmetric Yang-Mills }

To better understand the dynamical consequence of the
duality, it is necessary to have a control of the quantum properties of
the SYM theory. We have just shown that there exists a preferred choice of
variables for expressing the duality. The SYM theory is defined on a
flat Minkowski space with 
Lagrangian density \eq{s1}-\eq{s4}, the SUGRA metric is
given by 
%c9
\eq{brinkman3}. 
This choice of variables is an important simplification to allow for a
development of  the perturbation theory, which we will turn to now.

\subsection{A further rescaling and the SYM action}
 
For perturbative analysis, it is convenient to 
scale the fields further so that the kinetic terms are as close as
possible  to being canonically normalized and independent of the
coupling. To achieve this, we define
\be
A_{\mu} = \gym {\cal A}_{\mu}, \qquad 
Y^a = \gym Z^a, \qquad 
\psi = \gym \l. 
\ee
Then
\be
F_{\mu\nu} = \gym {\cal F}_{\mu\nu} + (\del_{\mu}\gym) {\cal A}_{\nu} -
(\del_{\nu}\gym) {\cal A}_{\mu},
\ee
where
\be
\cF_{\m\n} := \del_\m \cA_\n - \del_\n \cA_\m + i\gym [\cA_\m, \cA_\n]. 
\ee
Thus the YM action simplifies to
\be 
S_{\YM} = \int d^4 x \; \mbox{Tr} 
\Bigl[ 
\frac{-1}{4} \cF_{\m\n}\cF^{\m\n}
+\frac{\gym'}{\gym} \del_\m \cA^\m \cA_-
+  a \cA_-^2
\Bigr], 
\ee
where $a:= \frac{\gym'^{\, 2}}{2 \gym^2} -(\frac{\gym'}{\gym})' $.
To carry out perturbation analysis, one needs to fix a gauge. 
It turns out to be convenient to consider the following generalized
Lorentz gauge
\be
\del_\m \cA^\mu + f(x^+) \cA_- =0,
\ee
%c8 
which is a combination of the Lorentz gauge and the axial gauge $\cA_-=0$. 
%c8 This gauge fixing condition is invariant under the supersymmetry 
%c8 transformation for an arbitrary function $f(x^+)$
%p3 at the free field level.
Using the gauge fixing term
\be
S_{\rm g.f.} = \int d^4 x \; \Tr \Bigl[ -\frac{1}{2 \xi} (\del_\m \cA^\m+f(x^+)
\cA_-)^2\Bigr] ,
\ee
we have
\bea
S_{\YM} + S_{\rm g.f.} = \int d^4 x &\Tr  &\Bigl[
\frac{1}{2} \cA_\m \del^2 \cA^\m + \frac{1}{2}(1-\frac{1}{\xi}) 
%p1 (\del\cA)^2 
(\del_{\mu}A^{\mu})^2
+(\frac{\gym'}{\gym} -\frac{f}{\xi})
(\del_\m \cA^\m)
\cA_- +
(a-\frac{f^2}{2\xi})\cA_-^2 \Bigr]\nn\\
&&  \mbox{+ cubic and 
quartic 
terms}.
\eea
A particular simple gauge choice is therefore 
given by 
\be
\xi =1, \quad f= \gym'/\gym.
\ee
In this case, 
\be \label{mod1}
S_{\YM} + S_{\rm g.f.} = \int d^4 x \; \Tr \Bigl[
\frac{1}{2} \cA_\m \del^2 \cA^\m + \at \cA_-^2 \Bigr]  
\mbox{+ cubic and quatic terms},
\ee
where 
\be
\at : = -(\frac{\gym'}{\gym})'. 
\ee
 
Similarly, we obtain the scalar and fermion action
\bea  
S_X &=&
\int d^4 x \, \mbox{Tr} \Bigl[- \frac{1}{2}  {\cal
    D}^{\mu} Z^a {\cal D}_{\mu} Z^a 
+ \frac{\gym^2}{4} [Z^a, Z^b]^2 \Bigr],  \label{mod2} \\
S_{\Psi} &=& \frac
{1}
{2} \int d^4 x \, \mbox{Tr}\Bigl[ \bar{\l} \Gamma^{\mu}
  [-i{\cal D}_{\mu}, \l]  
+ \gym \bar{\l} \Gamma^a [ Z^a, \l] \Bigr], \label{mod3}
\eea
where 
\be
{\cal D}_{\mu} Z^a = \del_{\mu} Z^a + i\gym [{\cal A}_{\mu}, Z^a]
\quad \mbox{and} \quad 
{\cal D}_{\mu} \l = \del_{\mu} \l + i \gym [{\cal A}_{\mu}, \l]. 
\ee
 
Finally, in terms of the rescaled fields, the axionic coupling terms become
\bea
S_{\chi^B} &=& \int d^4 x \,  \gym^2 \chih' \, \Tr \Bigl[
\cA_-(\del_2 \cA_3 -\del_3 \cA_2) -\cA_2 \del_- \cA_3  
+i\gym \cA_- [\cA_2,\cA_3] \Bigr] , \label{mod4} \\
S_{\chi^F} &=& \int d^4 x \, \gym^2 \chih' \, \Tr \Bigl[ 
\frac
{-i}
{4} \bar{\l}  \Gamma^2\Gamma^3 \Gamma_- \l \Bigr], \label{mod5}
\eea
which give rise to correction to the propagators of $\l$ and $\cA_\m$ and 
a vertex involving $\cA_-, \cA_2,\cA_3$.

\subsection{Feynman rules}

In the following, we will consider the case when the SYM theory 
is defined for the whole line $-\infty< x^+ < \infty$. 
This may not be so when the SUGRA
background is singular. 
Later we will discuss the case 
when the SUGRA background has singularity (geodesic incompleteness) at  $x^+=0$.

The action given above (\ref{mod1}-\ref{mod5}) in terms 
of the fields $Z^a, \l, \cA_\m$ is
suitable for performing a perturbative analysis. We will 
treat the $\cA_-^2$ term in \eq{mod1}, \eq{mod4},
\eq{mod5} as perturbation. The propagators for the
scalar, the gauge boson and the Majorana fermions are
respectively
\bea
K_{ab} (x) &=&  
\int \frac{d^4 p}{(2\pi)^4}\frac{-i\d_{ab}}{ p^2}  e^{i p x},
\qquad a,b=4,\cdots,9, \label{Kab}\\
K_{\m\n} (x) &=&
\int \frac{d^4 p}{(2\pi)^4}\frac{-i \eta_{\m\n}}{ p^2} e^{i p x},
\qquad \m,\n = +,-,2,3,
\label{Kmn}\\
D(x) &=&  
\int \frac{d^4 p}{(2\pi)^4} \frac{i \ps}{p^2} e^{i p x}. \label{D}
\eea

Since the form of (\ref{mod1}-\ref{mod3}) is the same as 
the usual ${\cal N}=4$ SYM theory, 
the interaction vertices take the same form provided that 
one replaces the constant coupling with the
$x^+$-dependent one $\gym(x^+)$. In momentum space,  
the usual coupling constants $g_\a := \gym$, $\gym^2$ 
($\a=3,4$  for the 3-point and 4-point
vertices respectively) get replaced by 
\be
g_\a \d^{(4)} \left(\Sigma_I k_{I \m}\right) 
\rightarrow \gt_\a \d\left(\Sigma_I
 k_{I-}\right)\d^{(2)}\left(\Sigma_I k_{Ii}\right), 
\ee
where
\be
\gt_\a := \int dx^+ e^{i \sum_I k_{I+} x^+} g_\a(x^+),
\ee 
and 
\be
g_3(x^+):=\gym(x^+),\quad g_4(x^+) :=\gym^2(x^+)
\ee 
are 
defined for a 3-point vertex and a 4-point vertex respectively.  
We also note the following useful representation 
\be \label{g3}
\gt_3( k_{1+}, k_{2+}, k_{3+}) = \gym( -i\frac{\partial}{\partial k_{1+}})
\delta( k_{1+} +k_{2+} +k_{3+})
\ee
and
\be \label{g4}
\gt_4( k_{1+}, k_{2+},k_{3+}, k_{4+} ) = \gym^2
( -i\frac{\partial}{\partial k_{1+}} )
\delta(k_{1+} +k_{2+} +k_{3+}+k_{4+} ) .
\ee
As for \eq{mod4} and \eq{mod5}, 
the cubic term in the 
action \eq{mod4} gives rises to 
a new vertex involving $\cA_-, \cA_2,\cA_3$ with coupling $ i \gym^3
\chih'$. The rest of \eq{mod4}, \eq{mod5} 
\bea
\cL_2 &:=& \int dx \;  \gym^2 \chih' \Tr (\cA_- \del_2 \cA_3 -\cA_-\del_3 \cA_2
-\cA_2\del_- \cA_3),\\
\cL_3 &:=&  \int dx \; \frac
{-i}
{4} \gym^2 \chih' \Tr \lamb \G_2\G_3\G_- \l 
\eea
constitute corrections to the propagator.

\section{Wilsonian effective action: holographic reconstruction
  of the bulk metric}

We are interested in understanding the nature of spacetime singularity
from the dual gauge theory point of view. To do this, one needs to be
able to detect the properties of the bulk spacetime, in particular its 
$x^+$-dependence, 
from the gauge theory. The UV/IR relation \cite{SW,peet} is the key.
The relation gives a channel to probe the physics
in the interior of the bulk by looking at the dependence on $\L$ of SYM
quantities. According to it, 
introducing a momentum cutoff $\L$ in the SYM corresponds to 
bulk physics with a spatial IR cutoff at a 
certain value of the radial coordinate. 
This suggests to introduce a cutoff in the gauge theory. 
The question is which gauge theory quantity one should/could
use to probe or even reconstruct the bulk metric. 

%c6
In the approach of holographic renormalization group flow \cite{flow},
with certain regularity of the metric assumed, one can reconstruct the
bulk metric as a series expansion from the boundary out of the conformal
field theory data by solving the Einstein equation. This approach won't
be helpful for problems involving spacetime singularity, where the
regularity assumption is questionable. Also Einstein equation is
expected to be modified or break down completely. A new approach is
needed here.
 
In the following we will compute the 1-loop correction to
the quadratic fermion effective action and 
propose to use the 
UV/IR relation  to reproduce the metric of the bulk
from the boundary theory. 
We choose to look at the kinetic term of the fermion for simplicity. 
The kinetic terms of other fields will give the same information on 
the 
base space geometry due to supersymmetry.

\subsection{1-loop Wilsonian action: 
%c7 fermionic 
fermion kinetic term}

\begin{figure}[bpt] 
\begin{center}
\includegraphics[scale=.7]{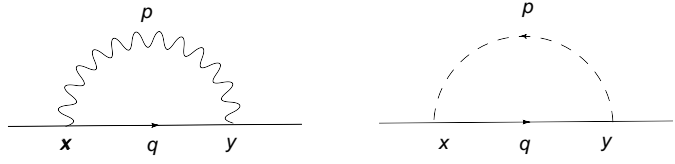}
\caption{One-loop contribution to the fermion kinetic term from 
gauge boson and scalars  
\label{Fig. 1}}
\end{center}
\end{figure}

For simplicity let us take $\chi=0$. 
The  fermion propagator receives 1-loop contribution from the Feynman
diagrams in figure 1 and figure 2. 
Both diagrams are planar.
For figure 1, summing over the contributions from the gauge and
scalar fields, we have 
\bea
I_1 &=&  N \int d^4 x d^4 y \Tr i \gym(x^+) \lamb(x) \Gamma^M D(x-y)\Gamma^N 
i \gym(y^+) \l(y)
K_{MN}(x-y) \nn\\
&=&   8 N \int d^4 x d^4 y \frac{d^4 p d^4 q}{(2 \pi)^8} \Tr \gym(x^+) 
\lamb(x) \frac{\qs}{p^2 q^2 }   \gym(y^+)\l(y) e^{i (p+q)(x-y)} \nn\\
&=& 4N \int d^4 x d^4 y \frac{d^4 p}{(2 \pi)^4} \Tr \gym(x^+) \lamb(x) 
F(p)  \ps  g(y^+) \l(y) e^{i p (x-y)}  ,
\eea
where we have  used $\Gamma^M \qs \Gamma_M = -8 \qs$ in the second step, 
performed a
change of variables $q\to q+p/2, p \to -q+p/2$ in the third step,
and introduced the definition
\be \label{F}
F(p) := \int \frac{d^4 q}{(2 \pi)^4} \frac{1}{(q-p/2)^2 (q+p/2)^2 }.
\ee
To simplify further, 
we note that 
\bea
&&\int d^4 x d^4 y \int \frac{d^4 p}{(2\pi)^4} f_1(x) F(p)\ps e^{ip(x-y)} f_2(y)
\nn\\
&& = \int d^4 x d^4 y f_1(x) \bigl(F(i\del_y) \cdot i\ds_y \d(x-y)\bigr) 
f_2(y)  \nn\\
&& = \int d^4 x f_1(x) F(-i\del)\cdot (-i \ds) f_2(x) 
+ \int d^4 x F(i \del) \cdot i\del_\mu \bigl(f_1 (x) \Gamma^\mu f_2 (x)\bigr),
\eea
where $f_1 = \gym(x^+) \lamb(x) $ and $f_2 =  \gym(x^+) \l(x)$
%c8
and we have performed an integration by parts in the last step.
The last term above 
%c8 (a boundary term)
vanishes since $\Gamma^0 \Gamma^\mu$ is symmetric in the Majorana representation.
Therefore we obtain
\be \label{I1-exact}
I_1 =  -4N \int d^4 x \Tr \gym(x^+) \lamb(x) F(-i \del)  i \ds 
\bigl( \gym(x^+) \l(x)\bigr) .
\ee 
Now the equation of motion has $i \ds \l =$ 
quadratic
in fields, therefore
$\ds \l$ can be treated as zero for the fermion kinetic term we are computing.
As a result, $I_1$ 
simplifies
to
\be \label{I1-eom}
I_1 = - 4N \int d^4 x \Tr \gym(x^+) \lamb(x) F(-i \del)  
\bigl( i \Gamma^+ \gym' \l(x)\bigr) .
\ee

\begin{figure}[bpt] 
\begin{center}
\includegraphics[scale=.7]{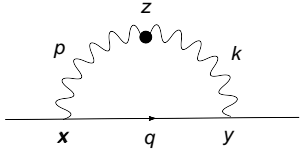}
\caption{Contribution due to propagator corrections  
\label{Fig. 2}}
\end{center}
\end{figure}  

The second source of contribution comes from the  $\cA_-^2$ vertex from
\eq{mod1}. There is only one diagram (figure 2) since only one such insertion 
can be made.  We have
\bea
I_2 &=& N \int d^4 x d^4 y d^4 z \Tr i \gym(x^+) \lamb(x) \Gamma^+ 
D(x-y) \Gamma^+ i \gym(y^+)\l(y) 
K_{+-} (x-z) 2 i \at(z^+) K_{+-} (z-y)\nn\\
&=& 4N\int d^4 x d^4 y d^4 z \frac{d^4 p d^4 q d^4 k}{(2\pi)^{12}} 
\Tr \gym(x^+) \lamb(x) \Gamma^+ \frac{q_-}{q^2 p^2 k^2} \gym(y^+)\l(y)
\at(z^+) e^{iq(x-y)} e^{ip(x-z)}e^{ik(z-y)}\nn\\
&=& 4N\int d^4 x d^4 y d^4 z \frac{d^4 p d^4 k}{(2\pi)^{12}} 
\Tr \gym(x^+) \lamb(x) \Gamma^+ G(p,k)\gym(y^+)\l(y)
\at(z^+) e^{ip(x-y)} e^{ik(z-y)},
\eea
where, in the last step, we have shifted the 
momenta 
as $q \to q+p/2, p \to -q+p/2, k\to
k-q+p/2$  and introduced the kernel
\be \label{G}
G(p,k) := \int \frac{d^4 q}{(2\pi)^4} \frac{(q+p/2)_-}{(q+p/2)^2 (q-p/2)^2
(k-q+p/2)^2}.
\ee 
By performing a similar manipulation like the one above for $I_1$, we finally 
arrive at
\be \label{I2-exact}
I_2 = 4 N \int d^4 y \Tr \Bigl( G(i\del_x, i \del_z) \gym(x^+) \lamb(x) \Gamma^+
\at(z^+)\Bigr)\Big|_{x=z=y} \gym(y^+) \l(y).
\ee
The effective action is given by the sum $I_1+ I_2$ and is governed by
the behaviour of the kernels $F(p), G(p,k)$ given in \eq{F} and \eq{G}. 

Now, to obtain the Wilsonian effective action, one would like to
integrate out 
oscillation
modes with momentum above a cutoff scale $\L$
and replace their contribution to low momentum modes by introducing new
interaction vertices in the effective action. However, although the
separation of modes is well defined for theory with only global
symmetries, the separation into low and high momentum modes does not
respect gauge symmetry and hence is not a well-defined procedure. The
subtleties concerned with the definition of Wilsonian action for gauge
theory were recently discussed in \cite{bilal}. Instead of using a
separation of momentum modes into low and high frequency ones, \cite{bilal}
proposes an alternative procedure by separating the loop momentum into
low and high region.   
In general, the Wilsonian action obtained in this manner contains
non-gauge invariant terms. Moreover since one can always shift the loop
momentum, one needs to give a specific prescription to avoid any
ambiguities. For the 1-loop case, the proposed prescription is, by
utilizing the Feynman parametrization, to first reduce the one-loop
integral into a certain standard form 
where there is no linear dependence in the loop momentum in the denominator.
Then a IR cutoff on the loop momentum is imposed on this
integral. And it has been shown that all non-gauge invariant terms
cancel in the case of supersymmetric gauge theories \cite{bilal}. 

We remark that in general one may use other prescription to impose a 
Wilsonian cutoff. This corresponds to different definitions of the
Wilsonian effective action which are equivalent in the following sense:
the cutoff $\L$ is introduced as an infrared cutoff when computing the
Wilsonian action. It becomes a UV cutoff when one uses the Wilsonian
action to compute correlation functions. Although one may obtain
different Wilsonian actions with different ways to impose the IR cutoff,
as long as one uses the same corresponding prescription for the
UV cutoff,  one will get the same 
correlation functions when using the Wilsonian action to compute correlators.

Now let us introduce the Wilsonian cutoff following the above
prescription. The  kernels with high loop momenta modes integrated out 
are given by
\be \label{F-cut}
F_W(p) := \int_0^1 d\a \int_\L^\infty  \frac{d^4 q}{(2 \pi)^4} 
\frac{1}{\bigl(q^2 +\a (1-\a) p^2\bigr)^2 },
\ee
\be \label{G-cut}
G_W(p,k) := 2 \int_0^1 d\a_1 d\a_2 d\a_3 
\int_\L^\infty \frac{d^4 q}{(2\pi)^4} 
\frac{\delta(\a_1+\a_2+\a_3-1)[(1-\a_1)p_- +\a_3 k_-]}
{\bigl(q^2 +\a_1(1-\a_1)p^2+ \a_3(1-\a_3)k^2+2\a_1\a_3 p\cdot
k\bigr)^3}.
\ee
As noted above, 
their contributions to low momentum modes then appear as new interaction
vertices in the Wilsonian action. 

The kernels $F_W, G_W$ can be evaluated and give  
an expansion of the Wilsonian 
action in derivatives of 
the field $\l$.  Let us first start with  $I_1$.
It is straightforward to compute $F_W$ and we have
\bea \label{F-exp}
F_W(p) &=& \frac{1}{16 \pi^2}\left[ 
C  +1 - \frac{4+2y}{ \sqrt{(4+y)y}}
\sinh^{-1}(\frac{\sqrt{y}}{2}) \right], \qquad y:= p^2/\L^2, \nn\\
&=& \frac{C}{16 \pi^2} +
\sum_{n=1}^\infty \frac{(n-1)!(n+1)!}{(2n+1)!}\left(\frac{-p^2}{\L^2}\right)^n, 
\eea
where $C = \log s|_{\L^2}^\infty$ is an infinite constant. This does not
contribute to \eq{I1-eom}. Using $-p^2 = \del^2 = -2 \del_- \del_+ + \del_i^2$,
it is easy to find 
\be
I_1 = \frac{i N}{6 \pi^2 \L^2} \int d^4 x \gym'^2 \Tr \lamb \Gamma^+ \del^+ \l 
+ \cdots,
\ee 
where $\cdots$ are terms of second or higher derivatives 
of
$\l$. 
We have kept only terms which are first order in derivatives of $\l$ since,
as we will see in the next subsection, 
these terms may be interpreted as due to a nonzero component  $g_{+ +}$ of the  
metric.

As for $I_2$, it is easy to evaluate the $q$-integral and get
\bea \label{G-exp}
G_W(p,k) 
&=&
\frac{1}{16\pi^2}\int_0^1 d\a_1 \int_0^{1-\a_1} d\a_3
[(1-\a_1)p_- + \a_3 k_-]\frac{\Delta+2\L^2}{(\Delta+\L^2)^2} \nn \\
&=&\frac{1}{8\pi^2}\left[
\frac{p_-}{3\L^2}-\frac{(6p^2+6p\cdot k+7k^2)p_-}{80\L^4}+\cdots\right],
\eea
where $\Delta = \a_1(1-\a_1)p^2+\a_3(1-\a_3)k^2+2\a_1\a_3 p\cdot k$,
as an momentum expansion. We have dropped the $k_-$ term in 
the last line above %\eq{G-cut}
since $k_-=0$ when acting on a function of $z^+$ 
and hence this term does not
contribute in \eq{I2-exact}. 
Substituting \eq{G-exp} into \eq{I2-exact},
we obtain
\be
I_2 = 
\frac{i N}{3 \pi^2 \L^2}
\int d^4 x \gym'^2 \Tr \lamb \Gamma^+ \del^+ \l + \cdots,
\ee
where, again, $\cdots$ denotes terms of second or higher derivatives of $\l$.

Concentrating on the first derivative terms, we find the 1-loop Wilsonian action 
$\Gamma_{\rm eff,1} = I_1+I_2$,
\be
\Gamma_{\rm eff,1} =
\frac{iN}{2\pi^2 \L^2} 
\int d^4 x \gym'^2 
 \Tr \lamb \g^+ \del^+ \l .
\ee
Substituting $\gym^2 = 4 \pi e^{\phi}$, 
we finally obtain 
\be \label{eff}
\Gamma_{\rm eff,1} \approx 
 i
\int d^4  x 
\frac{N e^\phi}{2\pi\L^2}
 {\phi'}^2  
\Tr  \lamb\Gamma^+ \del^+ \l .
\ee

\subsection{Holographic reconstruction of bulk metric}

Next we want to find an interpretation of the result \eq{eff} which
will allow us to reconstruct the bulk metric. We start by noting that
two different UV/IR relations have been considered
\cite{peet}. 
In terms of our coordinates, these are 
\bea
u &
\sim
&  \frac{1}{\L},\label{uvir1} \\
u &
\sim
& \frac{\gym N^{1/2}} {\L}. \label{uvir2}
\eea 
In the original $AdS_5 \times S^5$ case, 
the two relations are similar except for an overall constant which depends on the
gauge coupling. 
The holographic relation \eq{uvir1} corresponds to a probe by one of the
massless supergravity fields and can be derived by a 
scaling argument for the wave equation in the SUGRA
background. This relation has been applied to the counting of entropy
\cite{SW}. 
The holographic relation \eq{uvir2}  is relevant for the effective
action of a D3-brane probe at a distance $u$ and is derived by a 
stretched open string attached to the D3-brane.
  
In our case with a time dependent coupling $\gym$, the two relations are
distinctly different. 
If we employ the holographic relation
\eq{uvir2},
but more precisely:
\be \label{location}
u = \frac{\gym(x^+) N^{1/2} }{  \L} \frac{1}{\pi},
\ee
then the effective action \eq{eff} can be written as
\be \label{eff2}
\Gamma_{\rm eff,1} \approx  
i \int d^4  x \; \frac{u^2}{8}  {\phi'}^2
\Tr  \lamb\Gamma^+ \del^+ \l.
\ee  
Compared to \eq{uvir2}, 
eq.\eq{location} 
includes an additional numerical factor of
$\pi$. 
We remark that previous tests of the holographic
relation is not sensitive to the overall numerical factor. 
Here it is fixed
by requiring 
a matching with the bulk metric as we will demonstrate.

Next let us compute the kinetic term for the
fermion field for 
%c8 a 
D3 branes placed at $u$ of \eq{location}. 
Normally a D3 brane sitting at a constant $u$ is 1/2 BPS.
It is quite remarkable that for our supergravity background, 
a D3-brane sitting at an arbitrary $u=u(x^+)$ is 
also 1/2 BPS.
To see this, consider the action for a D3-brane probe with zero
worldvolume field strength, 
\be
I = \int d^4 x e^{-\phi} \sqrt{-\det G_{\m\n}} + \int C ,
\ee
where  
\be
G_{\m\n} =  \frac{\del X^M}{\del x^\m}
\frac{\del X^N}{\del x^\n}  g^{(s)}_{MN}
\ee
is the pull back to  D3-brane worldvolume of the 
spacetime metric in the string frame $g^{(s)}_{MN}= e^{\phi/2} g_{MN}$ 
and   $g_{MN}$ is given by 
%c9
\eq{brinkman3}. For a D3-brane 
in the static gauge $X^\mu = x^\m$, $\m=+,-,2,3$ and with $u=u(x^+)$, 
it is
\be
G_{\m\n} = \frac{e^{\phi/2}}{u^2}(\eta_{\m\n} + \frac{\Omega}{2} u^2
\d_{\m +} \d_{\n +}).
\ee
It is easy to check that the variation of the Born-Infeld term 
$\d I_0/\d u
$ cancels 
against the variation $\d I_{WZ}/\d u
$ of the WZ term. 
The equation of motion for $u$ is thus
\be
\del_\mu \frac{\d I_0}{\d( \del_\m u)} = 
\del_\m  \bigl(\frac{\sqrt{-G}}{u^2} (G^{-1})^{\m\n} \del_\n u
\bigr) =0,
\ee
which is satisfied for arbitrary $u(x^+)$ since $(G^{-1})^{++} =0$.
As for supersymmetry, the preserved supersymmetry
is given by the kappa-symmetry condition
\be
(1- \G) \e =0,
\ee
where 
\be
\G = \frac{-i}{4! \sqrt{-G}} \e^{\m_1 \cdots \m_4} \del_{\m_1}
X^{M_1}\cdots \del_{\m_4} X^{M_4} \Gamma'_{M_1 \cdots M_4},
\ee 
\be
\Gamma'_M = E_M^A \Gamma_A
\ee
and $\Gamma_A$ are the flat space $\Gamma$-matrices. 
$\e$  has to satisfy also the condition \eq{g-eps} of the \IIB supergravity
background. For our D3-brane, it is easy to obtain
\be
\G = -i (\G^{23} + \G^{23-+} - u' \G^{23r+}).
\ee
Using $\G^+ \e =0$, this reduces to $\G = -i \G^{23}$. Thus we conclude
that the D3-brane 
is supersymmetric for the SUSY parameter satisfying
the projector conditions \eq{g-eps} and $(1- i \G^{23}) \e =0$.

For such a D3 brane,  the bulk metric 
%c9 
\eq{brinkman3} at nonzero 
$u(x^+)$ gets an additional contribution
%c9 
%c9 component $g_{++}$ compared to the flat metric at $u=0$, 
%c9 and the 4D part of the bulk metric is (\ref{brinkman3})
%c9
%c9 also put hat over g_++ below. 
and the induced metric 
(apart from the factor  $R^2/u^2$) is 
\be \label{4Dmetric}
ds^2_{4D}  = -2 dx^+ d x^- + dx_i^2+ \left(\frac{1}{4}\phi'{}^2 
+ u'^2/u^2 \right) u^2 dx^+{}^2
:= \eta_{\mu\nu} dx^{\mu} dx^{\nu} + \gh_{++} dx^+{}^2.
\ee
The curved space 
gamma
matrices $\g_\m$ are related to the flat space ones by
\be
\gamma^- = \Gamma^- +\frac{1}{2}\gh_{++}\Gamma^+, \quad \gamma^+
=\Gamma^+,\quad \gamma_i =\Gamma_i.
\ee
Therefore among other terms, there will be a kinetic term for the
fermion:
\be
-\frac{i}{2} \bar{\lambda} \gamma^{\mu} \del_{\mu} \lambda 
= -\frac{i}{2} \bar{\lambda} \Gamma^a  \del_a \l
+\frac{i}{4}\gh_{++} \lamb \Gamma^+ \del^+ \lambda. 
\ee
%c7 Thus 
And we expect an additional term
\be \label{lowestorder}
\frac{i}{4} \int d^4 x \, \gh_{++} \Tr  \lamb\Gamma^+ \del^+ \l 
= i \int d^4 x \, \frac{u^2}{8} \left(
\frac{1}{2}\phi'^2 + 2 u'^2/u^2
\right)\Tr 
\lamb\Gamma^+ \del^+ \l 
\ee
in the kinetic action of $\l$ in addition to the kinetic term for flat space.
Using the UV/IR relation \eq{location}, this is precisely equal to \eq{eff2}.

Thus we observe that the fermion kinetic term in the one-loop Wilsonian
action seems to know about the bulk metric. 
A priori, the 1-loop correction may be more general than 
being equivalent to turning on the 
$\gh_{++}$ component of the metric.
In addition, the functional form of the 
$\gh_{++}$ component 
is precisely reproduced.
In general, the D3-brane probe action cannot be identified with 
the Wilsonian action at the scale (\ref{uvir2}). 
However, sometimes there is supersymmetry protecting certain loop
amplitudes \cite{peet}. In particular it seems to be the case for the
kinetic term of the Wilsonian effective action. 
This leads us to the
proposal to identify the metric of bulk spacetime from the kinetic term of
the Wilsonian action.

% p5be
More explicitly, we are proposing a relation between 
the metric derived from the Wilsonian action 
and the induced metric in the D3-brane probe action
\be g^{\rm (YM)}_{MN} (\lambda, g_s) = g^{\rm (Bulk)}_{MN} (\lambda, g_s), 
\qquad M,N= +,-,2,3,u.
\ee
In general, the metric on the left hand side is valid
only when 't Hooft's coupling is small, 
while the quantity on the right hand side is good only when
't Hooft's coupling is large. 
%c10 Our conjecture is that supersymmetry protects this relation
%c10 from nonperturbative effects.
% p5ee
Our conjecture is that this relation is protected by supersymmetry. 

As remarked above, 
the form of the Wilsonian action is generally dependent on the
scheme implementing the infrared cutoff. 
 An immediate problem arises if one 
would like 
%c8 (as we do) 
to propose it to be
in correspondence with bulk gravitational physics 
since the latter, at least the bulk geometry, should be independent of
any particular cutoff scheme. 
In the above, if we have shifted the loop momenta and then impose the cutoff,
this will lead to 
a different coefficient, in the $1/\L$ 
expansion, for the $p^2/\Lambda^2$ term of $F(p)$ 
and for the $p_-/\Lambda^2$ term of $G(p,k)$. This has the effect of
changing the overall coefficient of the effective action 
$\Gamma_{\rm eff,1}$ \eq{eff}.
However this will only result in a modification to the formula (\ref{location})
which matches $u$ with $\L$ by an overall constant.
Thus one can always reproduce the bulk metric from the Wilsonian effective action 
of SYM.

Strictly speaking, it is not clear to what extent the supersymmetry 
can protect loop correction in matching D3-probe action with
the Wilsonian effective action of the boundary SYM.
Let us recall that while 
the $v^4$ term of SUGRA scattering amplitude is correctly 
reproduced in field theory \cite{DKPS}, 
higher momentum dependence is not \cite{DT}. 
We showed above that the lowest derivative terms fixed by the bulk metric 
(\ref{lowestorder}) are correctly reproduced. 
In the effective action there are terms involving higher derivatives of 
$\l$. 
It is not difficult to include these higher order contributions.
It remains to be seen whether 
and how these terms
in the Wilsonian effective action 
correspond to
quantum effects on the gravity side.

A similar calculation may be performed at higher loops. 
For example, the metric component 
%c9 
$\gh_{++}$ we calculated above will be
modified. Generally it takes the form:
\be \label{gh}
\gh_{++} = a(\phi,u^2) (\phi')^2,
\ee
where we have used the UV/IR relation to replace the $1/\L^2$ depenence
with $u$-dependence. 
%c9
Higher loop amplitudes contribute higher order in powers of $\gym$, but
only a $(\phi')^2$ dependence.
%c9 For example, $H_2 \sim {\cal O}(N^2 \gym^4)$. 
In general, following the above procedue, the 
higher loop corrections in gauge theory gives a 5d metric:
\be \label{quantum-metric}
%c10 
ds^2 = \frac{R^2}{u^2}( g^{\rm (YM)}_{\m\n}(u,x^+) dx^\m dx^\n +du^2). 
\ee
%c9
Due to the complicated $u$-dependence, 
the metric will generally not satisfy the Einstein equation. 
%c9 One may interpret the 
The dual  supersymmetric Yang-Mills provides a framework for
computing the quantum corrections to the supergravity action. 
This includes higher derivatives corrections in general. 
It will be very interesting to see if one is able to reproduce 
some of the well-known result, 
e.g. the $R^4$ term \cite{R4},  
from the SYM theory.

\subsection{On resolution of spacetime singularity }

The above analysis is performed for a correspondence which is defined
over the whole real line of $x^+$. In case when the SUGRA background is
singular, say at $x^+=0$, %and is defined only on the half line $x^+>0$, 
the SUGRA solution in the two regions $x^+>0$ and $x^+<0$ are actually 
two different solutions
as, at least classically, the degrees of freedom don't talk to each other.
A possibility is that the SUGRA solution restricted to, say $x^+>0$,
should be matched with the dual SYM constructed on $x^+ >0$
\footnote{
The Feynman rules will become more complicated. 
Principal value appears in \eq{g3} and \eq{g4} in 
addition to the Dirac delta function.
}.
However, it is also possible that stringy or quantum corrections to the 
SUGRA solution 
will resolve the singularity,
and both regions must be included in the complete theory.
The dual SYM theory will then be defined on the whole real line of $x^+$.
In particular, when the SUGRA singularity
corresponds to the vanishing of the YM coupling at $x^+ = 0$, 
the dual SYM theory is weakly coupled around the point $x^+ = 0$,
and has no reason to break down.
What happens in the bulk must be that 
the stringy and/or quantum corrections resolve the SUGRA singularity 

We will now argue that the second possibility is the generic 
scenario,
whenever the dual SYM theory is well defined.
Despite the fact the two regions are separated by the singularity, 
one may still put 
the two halves of the SUGRA background together and
consider the total theory $S = S_1 + S_2$, where 
$S_1$ or $S_2$ describes the SUGRA on $x^+>0$ or $x^+<0$ respectively. 
The dual SYM is now defined on the whole $x^+$ and the Feynman
rules take on the simple form as described before. In this description, 
SUGRA restricted to one of the two regions can be described as a subsector 
of the SYM
theory. 

%c9 
Classically, $S_1$ and $S_2$ don't interact with each other. However, in the SYM,
the replacement of principal value by delta function in the Feynman rules 
means something nontrivial must happen.  From the result of the previous subsection,
we see that the singular metric is reproduced by the SYM at the 1-loop
level.  Taking into account of  the corrections 
from higher loops, the dual bulk metric becomes \eq{quantum-metric}.
It is possible that the  metric becomes regular 
after taking into account of these higher order
corrections.
%c9 joint paragraph
A rather generic argument supports that spacetime singularity is indeed resolved.
In the Wilsonian effective action, since oscillations of frequency above 
an energy scale $\Lambda$ are integrated out, 
the higher derivative correction to the kinetic term must be such that 
the background geometry is smeared over a length scale of $\Delta x \sim 1/\L$. 
A potential singularity in $g_{++}$ is thus always resolved for any finite $\L$.
It will be very interesting to check explicitly if this is really the case.

\section{Discussions}

In our perturbative analysis above we have turned off the 
axion 
coupling ($\chi' = 0$) for simplicity.
Since both type \IIB SUGRA and ${\cal N} = 4$ SYM have 
the $SL(2, \mathbb{Z})$ symmetry \cite{SL2Zsymmetry} which 
mixes the dilaton and axion fields
\be
\tau = \chi + ie^{-\phi} \rightarrow 
\tau' = \frac{a \tau + b}{c \tau + d},
\ee
the extension of our results to a nontrivial axion field background 
should presumably be a direct result of the $SL(2, \mathbb{Z})$ symmetry, 
assuming that there is no technical difficulty in 
manifestly preserving this symmetry, 
as well as gauge symmetry and supersymmetry
in the Wilsonian effective action.

We have shown in this paper that 
the spacetime metric can be holographically reconstructed
from the kinetic term of the SYM Wilsonian effective action in 
the leading order approximation. Higher order
corrections to the kinetic term of
the Wilsonian action include quantum corrections 
(both $\a'$ and $g_s$) to
the SUGRA equations of motion.  
Due to the nature of the Wilsonian action where high momentum modes
are integrated out, it is expected that singularity in the metric will be
resolved. The confirmation of this will be very interesting.

We remark that in the  matrix cosmology proposal 
\cite{csv,miao}, it has been suggested that as one approaches the 
singularity,  the classical picture of spacetime breaks down since  
the non-diagonal 
degrees
of freedom of the matrix model get lighter and lighter 
and their effects cannot be ignored, and  that the singularity could be resolved by 
including these light modes in the description. Doing so, spacetime is
replaced by nonabelian matrices.
Our proposal is different. 
In our proposal, we have suggested a mechanism how 
spacetime singularity could
be resolved by including 
%c9 
all the 
quantum corrections to the Einstein equation, which in principle could be
computed from the gauge theory. After the resolution, ordinary spacetime
is still a valid concept.
However, without understanding the nature and organization principle of these
corrections, one does not actually feel one has 
a good
understanding of the 
physics involved. It is usually 
believed
that some form of quantized
spacetime and noncommutative geometry will be relevant at very
small distance scale. In the case of noncommutative quantum field
theory, one way to think about the noncommutative geometry is that it is
an effective and geometrical way to encode the Moyal phase factor.
It may be possible that 
%c9 such 
the infinite series of quantum 
corrections follows from 
some form of underlying noncommutative geometry.  
%c9 
If this really happens, the noncommutative geometry description will be
a better one than the classical spacetime. 
It is interesting to explore this possibility.

%%%%%%%%%%%%%%%%%%%%%%%%%%%%%%%%%%%%%%%%%%%%%%%%%%%

\section*{Acknowledgements}

We thank Ed Corrigan, Nick Dorey, Harald Dorn,
Chris Fewster, Kazuyuki Furuuchi, Jerome Gauntlett, 
Hsien-Chung Kao, James Liu, Kelly Stelle, 
Takeo Inami, Shunsuke Teraguchi,
Wen-Yu Wen, Toby Wiseman, Tamiaki Yoneya and Syoji Zeze 
for valuable discussions.
CSC also thanks the Isaac Newton Institute for Mathematical Sciences 
for hospitality,
and participants of the programme "Strong Fields,
Integrability and Strings" for discussions. 
The work of CSC is supported in part by EPSRC and PPARC.
The work of PMH is supported in part by
the National Science Council, and the National Center for Theoretical
Sciences, Taiwan, R.O.C.

%%%%%     %%%%%     %%%%%     %%%%%     %%%%%     %%%%%     %%%%%     %%%%%    
%%%%%     %%%%%     %%%%%     %%%%%     %%%%%     %%%%%     %%%%%     %%%%%

\end{document}